# Optimization research for rescue hot standby EMU location and coverage area in a large-scale high-speed railway network

Boliang Lin[a,*], Zhenyu Wang[a], Yaoming Shen[a], Yufei Meng[a]

[a] School of Traffic and Transportation，Beijing Jiaotong University，Beijing 100044, China

**Abstract:** With the extension of China high-speed railway network, the number of railway stations is increasing. The challenge that railway companies are facing is how to reasonably plan the location of hot standby Electric Multiple Units (EMU) and determine the rescue coverage area of each hot standby EMU. It is uneconomical to configure a hot standby EMU for each station. Railway companies want to use the minimum number of hot standby EMUs to provide rescue tasks for the entire high-speed railway network. In this paper, we develop the optimization models for hot standby EMU location and coverage area, respectively. The models aim to maximize the rescue distance and minimize the number of hot standby EMUs. What's more, in order to reduce the complexity of the models, a method of merging railway lines is used to transform the "point-to-network" to "point-to-point" rescue problem. Two numerical examples are carried to demonstrate the effectiveness of the models. The developed models are linear, and we use Python 3.7 to call Gurobi 9.5.2 to solve the models. In the results, we can find that the developed models can provide an optimal and reasonable plan for the location and coverage area of hot standby EMUs.

## 1. Introduction

In China, the EMU can be classified into three types: hot standby EMU, operational EMU, and under-repair EMU (maintenance condition). The hot standby EMU, by definition, refers to EMUs that have been repaired, are used as rescue EMUs, and can carry out rescue tasks at any time. When there is an accident on a running operational EMU, the hot standby EMU will carry out the rescue task and depart from the rescue station to accident site. According to the regulations on the use of EMU, there are three scenarios where hot standby EMU is required to carry out rescue task:

(1) When an operational EMU has been in operation for a certain period of time or distance, maintenance operations need to be carried out. In this case, the hot standby EMU is required to replace the operational EMU and carry out transportation tasks.

(2) An accident occurs on a running EMU, it is necessary to replace the running EMU with a hot standby EMU to complete transportation tasks.

(3) If there is a significant delay affecting a running EMU, specifically when the delay time exceeds an acceptable limit, resulting in unable to complete transportation tasks within a reasonable time frame, it is necessary to use a hot standby EMU to replace the delayed EMU and carry out the transportation tasks.

Hot standby EMU is usually stationed at a designated rescue station, and a rescue station is usually equipped with a hot standby EMU. The hot standby EMU location and coverage area problems are to determine which station is equipped with a hot standby EMU and determine its coverage area. In this way, this problem can also be named as rescue station location and coverage area problem. That is, determine which station is rescue station (equipped with a hot standby EMU). In a high-speed railway network, different rescue stations are responsible for rescue tasks of different areas. And each rescue station has its own coverage area, which means that if a running EMU experiences an accident within the coverage area of a rescue station, the hot standby EMU of the rescue station will depart from the rescue station to the accident site.

The cost of producing a hot standby EMU is expensive, making it uneconomical to equip every station with one. If the location of rescue stations is overly dense, it will lead to an increase in the production cost of hot standby EMUs. Conversely, if the location is too sparse, hot standby EMUs may not be able to carry out the rescue tasks for some long-distance stations. Therefore, it is significant to plan and locate rescue stations equipped with hot standby EMUs to ensure an efficient coverage area, which motivates us to conduct this research.

Furthermore, with the extension of China high-speed railway network, the number of stations in the network is also increasing. To improve the efficiency of the use of hot standby EMUs, it is also necessary to optimize the location and coverage area of rescue stations, which can help railway companies minimize the costs of producing hot standby EMUs. And it is also another motivation for this study.

When a hot standby EMU arrives at the accident site, it may be tasked with either carrying out the accident EMU's circulation plan, transporting passengers, or towing the accident EMU. In this paper, we focus solely on studying the location and coverage area of hot standby EMUs. For which rescue tasks carried out by hot standby EMUs when they arrive at the rescue site, we think it will be another interesting research work for future research.

The remaining sections of this paper are as follows: Section 2 provides an overview of relevant literature and our contributions. Section 3 presents the description of hot standby EMU location and coverage area problems. In Section 4, we formulate two models for hot standby EMU coverage area and hot standby EMU location, respectively. Furthermore, we conduct numerical experiments in Section 5 to verify the effectiveness of the models. And the conclusion of this paper is presented in Section 6.

## 2. Literature Review

According to the literature we have searched, there are few studies on the location of hot standby EMU. This means this problem is relatively under-examined and insufficiently explored within the existing literature. The similar problems to hot standby EMU location are location problem and emergency resource location problem, which can provide us with some ideas to solve the problem. As a typical problem in operational research, there are many studies about location problem. Weber [1] developed the theory of industry location, he studied the problem of locating a warehouse on a plane, which aims to minimize the total distance between the warehouse and multiple customers (known as the Weber problem). In the work of Church [2], he improved the model based on the work of ReVelle and Swain [3], which introduced a new formulation for location problem, namely COBRA. Dantrakul et al. [4] used a p-median method to solve the facility location problem. A facility location model with bounds for the number of opened facility was developed, which aims to minimize the setup cost and transportation cost. In the work of An et al. [5], they developed a set of two-stage robust mathematical models, which aim to design the reliable facility location networks. By analyzing the characteristics of the models, they also designed a column-and-constraint generation method to solve the models. Albareda et al. [6] constructed a mathematical optimization model for discrete facility location problem, in which they developed two mathematical formulations for the problem, and a method based on a network flow theory was used to solve the model.

Furthermore, a sub-problem of location problem is to determine the coverage area of locations. Pirkul and Schilling [7] studied the maximal covering location problem. In their work, they considered the workload limits of locations, and an efficient method was developed to solve the established model. In the work of Berman and Krass [8], they further extended the covering location problem, and introduced a non-increasing step function to describe the coverage degree. Some mathematical formulations are developed to solve the problem. Davari et al. [9] introduced the fuzzy theory to maximal covering location problem and the travel time between two nodes was described as a fuzzy variable. A simulated annealing algorithm was designed to solve the established model. Coco et al. [10] used a matrix to describe the min-max regret maximal covering location problem, a mathematical formulation was developed. To verify the effectiveness of the model, a 2-approximation algorithm was designed and some numerical experiments were tested. Toregas et al. [11] regarded the emergency facility location as a covering problem of multiple nodes while considering the equal cost of the objective. They developed a linear programming model to solve the problem. Based on the previous work, there are many variants of location problem: arc-covering formation problem (ReVelle et al. [12]), multi-level location set covering model (Church and Gerrard [13]), using Stackelberg game to solve the competitive set covering problem (Hemmati and Smith [14]), using a minimum number of facilities to completely cover a region (Murray and Wei [15]), planning public postal network (Šarac et al. [16]), gas detectors location (Vianna [17]), unmanned aerial vehicle service coverage (Park et al. [18]).

The hot standby EMU location problem can also be studied as an emergency resource location problem. Emergency resource location problem, which is also a typical problem in operational research, is to determine the location and coverage of rescue facilities while considering the availability, frequency of calls (demand), maximum rescue distance, and response time. Rajagopalan et al. [19] studied the emergency medical service coverage problem, a mathematical optimization model, which is to minimize the number of ambulances and determine their locations, was developed to cope with changes in demand patterns. Cheng and Liang [20] studied the emergency rescue location problem for urban ambulance and railway emergency systems. A model with fuzzy multi-objectives was designed, and a generic algorithm was used to solve the model. What's more, a case study based on Taiwan railway system was conducted to verify the effectiveness of the model. In the work of Boonmee et al. [21], they studied the facility location problem which is related to emergency humanitarian logistics. By conducting a survey about deterministic facility location problem, dynamic facility location problem, stochastic facility location problem, and robust facility location problem, they evaluated the solution methods for each problem and carried case studies based on them.

If an accident occurs in the railway network, the rescue trains need to arrive at the accident site as soon as possible. And the location of rescue trains focuses on minimizing the total travel time and maximizing the total coverage (Bababeik et al. [22]). The optimization of rescue train location is to find the minimum number of emergency rescue trains which can completely cover the entire railway network, and can also enhance the resilience of the network (Tang and Sun [23], Wang et al. [24]). In addition, there are many scenarios similar to rescue train location, such as: the rescue logistics problem under uncertain environments (Chang et al. [25], Zhang et al. [26], Wang et al. [27]), maritime emergency rescue (Razi and Karatas [28], Karatas [29]).

As summarized above, there are many studies about emergency resource location problem. In railway system, many scholars often focus on the location of traditional rescue trains, and that of hot standby EMU location is few. This further motivates us to conduct this study. The contributions of this paper are as follows:

(1) A merge strategy is introduced to reduce the complexity of the problem, transforming the "point-to-network" rescue problem into a "point-to-point" one.

(2) Developing two mathematical models for hot standby EMU location and coverage area, respectively.

(3) Two numerical experiments, based on the designed high-speed railway network, are conducted to verify the effectiveness of the models.

## 3. Problem description

### 3.1 Hot standby EMU coverage area problem

The hot standby EMU coverage area problem is to determine the rescue area of each rescue station. The rescue area not only includes stations, but also railway lines. And the combined rescue areas of all rescue stations should effectively cover the entire high-speed railway network.

When an accident occurs on a running EMU, the hot standby EMU must promptly leave the rescue station for the accident site, carrying out the rescue task. As for the rescue station, it is the station equipped with a hot standby EMU. To solve the hot standby EMU coverage area problem, we can design a decision variable to signify the core of this problem: rescue station $i$ is responsible for the rescue task at site $j$. In fact, accidents in the railway system might happen at unpredictable locations, such as railway stations and lines. It is unrealistic if we divide the railway network into units of "1cm or 1m" when studying the problem. Because this will greatly increase the complexity of the problem, and result in an infinite number of variables in the mathematical model.

In this paper, we use a specific strategy to merge railway lines and stations. The strategy is described as: use the midpoint of the railway line between two adjacent stations as a boundary point. Lines to the left of this boundary point are merged with the left station, while lines to the right of it are merged with the right station. If a running EMU experiences an accident on the left of the boundary point, we consider it as occurring at the left station. Conversely, if the accident happens on the right of the boundary point, we think it occurred at the right station.

Taking Figure 1(a) as an example, there are four stations (1, 2, 3, 4) in Figure 1(a), and $i$ is the midpoint of link 1-2, $j$ is the midpoint of link 2-3, $k$ is the midpoint of link 3-4. According to the above strategy, if a running EMU experiences an accident on link 1-$i$, we think the accident occurs at station 1. Similarly, if a running EMU experiences an accident on link $i$-$j$, we think the accident occurs at station 2. If an accident occurs on link $j$-$k$, we think the accident occurs at station 3. If an accident occurs on link $k$-4, we think the accident occurs at station 4. In this way, the complexity of the problem can be greatly reduced. And we can transform the rescue operations of hot standby EMUs within the railway network from "rescue station-network" to "rescue station-station", that is, from "point-to-network" to "point-to-point".

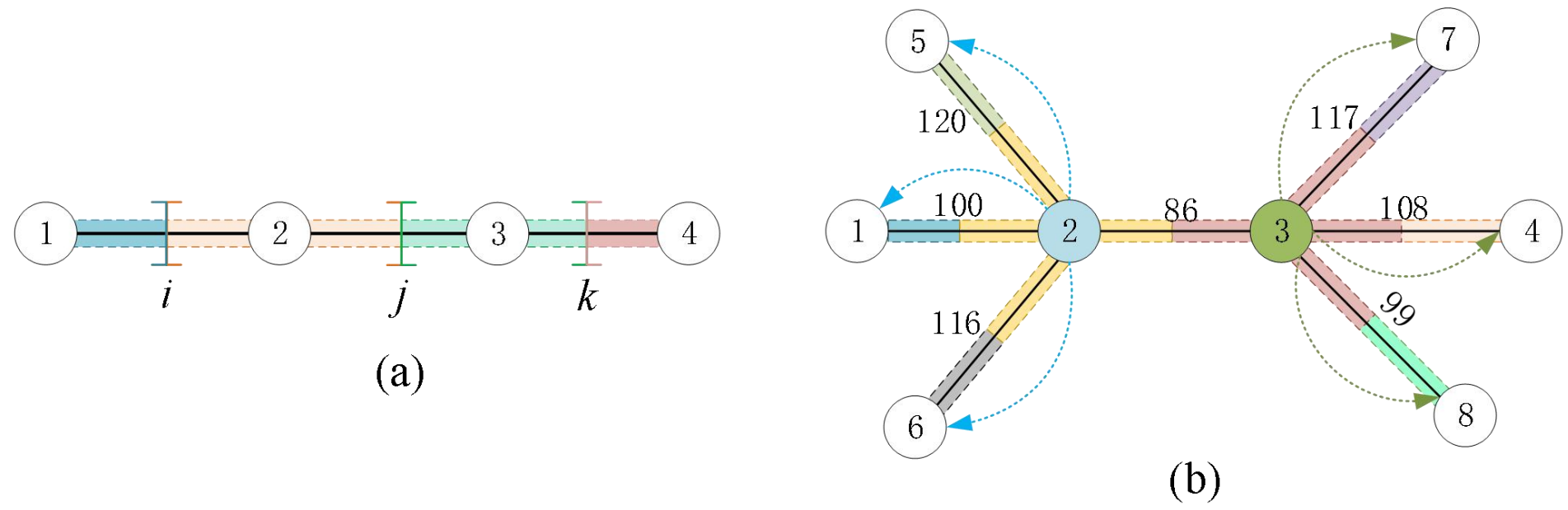

**Figure 1.** A simple high-speed railway network

Additionally, taking the high-speed railway network in Figure 1(b) as an example to illustrate the hot standby EMU coverage area problem. In figure 1(b), the numbered circles represent stations, and the numbers on the lines connecting stations are distance, unit: km. For example, the distance from station 1 to station 2 is 100 km. The colored links surrounding the stations are the affiliated areas of stations. For example, if a running EMU experiences an accident on the yellow links, we think the accident occurs at station 2.

What's more, stations 2 and 3 are rescue stations that equipped with hot standby EMUs. The dashed lines with directional arrows in figure 1(b) represent the rescue stations responsible for rescue tasks of different stations. In figure 1(b), rescue station 2 is responsible for the rescue tasks of stations 1, 2, 5, and 6. And rescue station 3 is responsible for the rescue tasks of stations 3, 4, 7, and 8. If a station itself is a rescue station, then it of course takes on its own rescue task.

It is important to highlight that if we adopt the strategy of merging railway network, the distances between stations need to be updated. The updated distance between rescue station $i$ and another station $j$ consists of two parts: 1) the original distance from station $i$ to station $j$, 2) half of the maximum value among the distances from station $j$ to its adjacent stations, while doesn't include the stations that in the route from $j$ to $i$. We also take the network in Figure 1(b) as an example, assuming that station 2 is responsible for the rescue task of station 3 (In fact, stations 2 and 3 are both rescue stations, and they of course carry out their own rescue tasks. Here, it is just to provide an example of updating the distance), the distance between stations 2 and 3 is updated as: $L_{23}+\frac{1}{2}\max\{L_{34}, L_{37}, L_{38}\}$, and it doesn't include $L_{32}$. In this way, station 2 serves all affiliated areas of station 3.

The coverage area of hot standby EMU, also called the coverage area of rescue station, should consider the rescue distance and rescue time of hot standby EMU. If a rescue station is far from the accident site, it may be unable to complete the rescue task and even affect the subsequent EMU operation on the line. Therefore, the core of optimizing the coverage area of hot standby EMU can be summarized as: which hot standby EMU (rescue station) is responsible for the rescue task of which station, while considering some practical constraints, such as the maximum rescue distance and maximum rescue time of hot standby EMU.

### 3.2 Hot standby EMU location problem

The location problem of hot standby EMU is a further extension of the coverage area problem. In the hot standby EMU coverage area problem, the rescue stations are known, that is, which stations equipped with hot standby EMUs are pre-given.

As the high-speed railway network expands, the rise in the number of stations necessitates an increase in rescue stations to cover the entire network. It is uneconomical and unscientific to equip each potential station with a hot standby EMU. How to reasonably and scientifically select the stations to be equipped with hot standby EMUs among potential stations and determine their rescue coverage areas is another challenge that needs to be solved.

Expanding on Figure 1 (b), Figure 2 represents an extended high-speed railway network. In figure 2, stations numbered 9 through 17 have been added. Among the newly built stations, the potential stations that could be equipped with hot standby EMUs are: 10, 11, 12, 14, 16, 17 (marked in orange circles). The location optimization problem of hot standby EMU is to determine which stations should be chosen as rescue stations, equipped with hot standby EMUs, from among the potential stations. As well as determining the coverage area of these rescue stations.

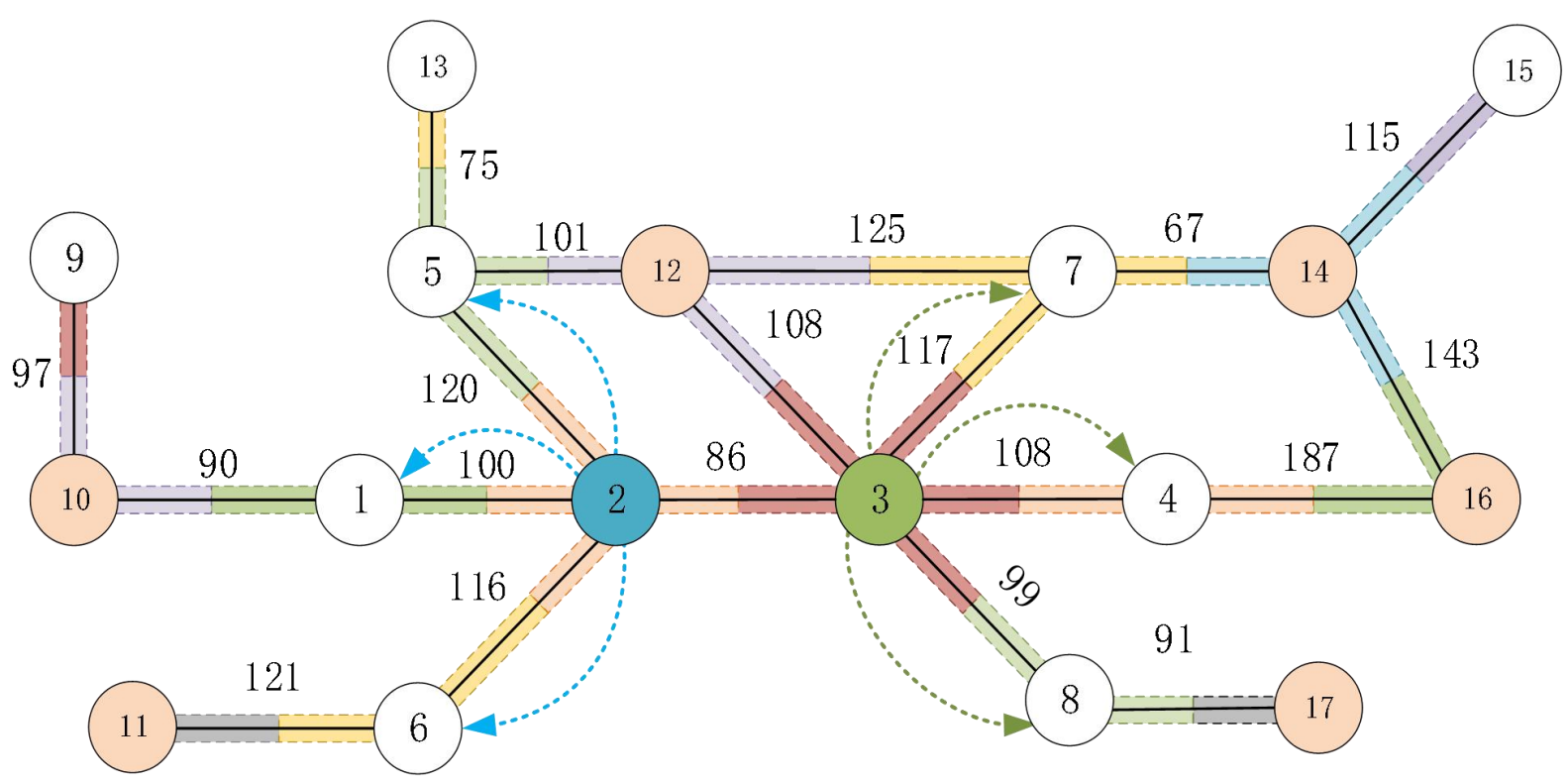

**Figure 2.** A simple railway network

Simultaneously, it should be noted that when optimizing the location of hot standby EMU for the high-speed railway network in Figure 2, existing rescue stations 2 and 3 should also participate in the optimization, and be considered pre-given rescue stations during the optimization. Although stations 2 and 3 are known rescue stations, due to changes in the high-speed railway network structure, the coverage area of them may also change. This is to achieve an optimal location for the entire network.

## 4. Mathematical Model

### 4.1 Coverage area optimization model for hot standby EMU

This section develops the hot standby EMU coverage area optimization model, while considering some constraints, including rescue distance, rescue time and so on. The symbols used in the model are shown in Table 1.

**Table 1.** Symbols used in the paper

| Symbol | Definition |
|---|---|
| $S^{\text{Station}}$ | Set of stations in a high-speed railway network. |
| $S^{\text{Depot}}$ | Set of rescue stations which are equipped with hot standby EMUs, $S^{\text{Depot}} \subset S^{\text{Station}}$. |
| $G(i,j)$ | Set of stations that are adjacent to station $j$ and doesn't include the stations that in the route from $j$ to $i$. Its detailed description is mentioned in Section 3.1. |
| $L^{\text{rescue}}$ | Maximum distance that a hot standby EMU can operate from rescue station to accident site, km. |
| $T^{\text{rescue}}$ | Maximum time that a hot standby EMU can operate from rescue station to accident site, hour. |
| $L_{ij}$ | The distance from station $i$ to station $j$, km. |
| $p_j$ | The probability of accidents occurring at station $j$, which can be determined by analyzing historical data and the annual accident frequency. |
| $N^{\text{HotEMU}}$ | The number of hot standby EMUs in a high-speed railway network. |
| $\mathbb{L}^{\text{Network}}$ | Total mileage of a high-speed railway network, km. |
| $Q^{\text{Workload}}$ | The maximum workload allowed for a rescue station (a hot standby EMU), and it is determined by $\mathbb{L}^{\text{Network}}$ and $N^{\text{HotEMU}}$, km. |
| $v$ | Average speed of a hot standby EMU, km/h. |
| $x_{ij}$ | Rescue coverage area variable. If rescue station $i$ is responsible for the rescue task of station $j$, the value is 1; Otherwise, the value is 0. |

The mathematical model aims to maximize the total rescue distance of hot standby EMUs, while considering the constraints of rescue distance and rescue time, as well as the maximum workload carried by hot standby EMUs. The coverage area optimization model for hot standby EMU (CMHSE) is as follows:

$$\max\ Z = \sum_{i \in S^{\text{Depot}}} \sum_{j \in S^{\text{Station}}} (L_{ij} + \frac{1}{2} l_{ij}^{\max}) x_{ij} \tag{1}$$

(CMHSE) s.t.

$$\sum_{i \in S^{\text{Depot}}} x_{ij} = 1, \forall j \in S^{\text{Station}} \tag{2}$$

$$(L_{ij} + \frac{1}{2} l_{ij}^{\max}) x_{ij} \le L^{\text{rescue}}, \forall i \in S^{\text{Depot}}, j \in S^{\text{Station}} \tag{3}$$

$$l_{ij}^{\max} = \max_{h \in G(i,j)} \{L_{jh}\}, \forall i, j \in S^{\text{Station}} \tag{4}$$

$$\tau_{ij} x_{ij} \le T^{\text{rescue}}, \forall i \in S^{\text{Depot}}, j \in S^{\text{Station}} \tag{5}$$

$$\sum_{j \in S^{\text{Station}}} (L_{ij} + \frac{1}{2} l_{ij}^{\max}) p_j x_{ij} \le Q^{\text{Workload}}, \forall i \in S^{\text{Depot}} \tag{6}$$

$$x_{ij} \in \{0,1\}, \forall i \in S^{\text{Depot}}, j \in S^{\text{Station}} \tag{7}$$

The objective function of CMHSE is to maximize the total rescue distance of hot standby EMUs. Constraint (2) ensures that each station is covered by a rescue station. Constraints (3) and (4) limit the running distance from rescue station to accident station. In constraint (3), we use the updated distance from station $i$ to station $j$ to represent the rescue distance, the updated distance consists of two parts: the original distance from station $i$ to station $j$, and half of the maximum value among the distances from station $j$ to its adjacent stations which are from $G(i, j)$. In this way, if rescue station $i$ carries out the rescue task of station $j$, it can also cover the surrounding area of station $j$. This is also the reason why we adopt the merging strategy, which can fully cover the entire railway network.

Constraint (5) limits the running time from rescue station to accident station, $\tau_{ij}$ is the average running time from rescue station $i$ to accident station $j$, and it is determined by $\tau_{ij} = (L_{ij} + 0.5 l_{ij}^{\max}) / v$. Constraint (6) shows that the workload of a rescue station can not exceed the maximum limitation. In practical railway operation, a rescue station is equipped with a hot standby EMU. The maximum workload permitted for a rescue station (a hot standby EMU) refers to the average service coverage area for each hot standby EMU across the high-speed railway network:

$$Q^{\text{Workload}} = \frac{1}{2} \times \frac{\gamma \mathbb{L}^{\text{Network}}}{N^{\text{HotEMU}}} \tag{8}$$

Where, $\gamma$ is the workload fluctuation coefficient. Constraint (6) only shows the one-way distance of a hot standby EMU from rescue station to accident site, and there is also a return distance when the hot standby EMU comes back to the rescue station. This is the reason for multiplying by 1/2 in constraint (8). And the calculation formula for $N^{\text{HotEMU}}$ is as follows:

$$N^{\text{HotEMU}} = \alpha N^{\text{EMU}} \tag{9}$$

In equation (9), $N^{\text{EMU}}$ is the total number of EMUs in a high-speed railway network and $\alpha$ is the proportion coefficient of hot standby EMUs.

**4.2 Location and coverage area optimization model for hot standby EMU**

CMHSE is to determine which station is serviced by which rescue station, and all rescue stations are pre-given. As the high-speed railway network expands, the number of railway stations will further increase, which means that the existing location of rescue stations will not be able to cover the entire high-speed railway network. Therefore, it is necessary to increase the number of rescue stations, which by choosing new rescue stations from the potential stations.

Here, we design a new variable to determine whether a station is equipped with a hot standby EMU:

$$y_k = \begin{cases} 1 & \textit{if} \text{ station } k \text{ is equipped with a hot standby EMU} \\ 0 & o.w. \end{cases} \tag{10}$$

We introduce some new symbols based on table 1: $S^{\text{Set}}$ is the set of potential stations that can be equipped with hot standby EMUs, it contains the stations that have already been equipped with hot standby EMUs, and there is $S^{\text{Depot}} \subset S^{\text{Set}}$. $c_k$ is the cost of equipping a hot standby EMU at station $k$, its unit is yuan (RMB). The location and coverage area optimization model for hot standby EMU (LCHSE) can be described as:

$$\max\ Z_1 = \sum_{i \in S^{\text{Set}}} \sum_{j \in S^{\text{Station}}} (L_{ij} + \frac{1}{2} l_{ij}^{\max}) x_{ij} \tag{11}$$

$$\min\ Z_2 = \sum_{k \in S^{\text{Set}}} c_k y_k \tag{12}$$

(LCHSE) s.t.

$$\sum_{i \in S^{\text{Set}}} x_{ij} = 1, \forall j \in S^{\text{Station}} \tag{13}$$

$$y_k = 1, \forall k \in S^{\text{Depot}} \tag{14}$$

$$x_{ij} \le y_i, \forall i \in S^{\text{Set}}, j \in S^{\text{Station}} \tag{15}$$

$$(L_{ij} + \frac{1}{2} l_{ij}^{\max}) x_{ij} \le L^{\text{rescue}}, \forall i \in S^{\text{Set}}, j \in S^{\text{Station}} \tag{16}$$

$$l_{ij}^{\max} = \max_{h \in G(i,j)} \{L_{jh}\} \quad \forall i, j \in S^{\text{Station}} \tag{17}$$

$$\tau_{ij} x_{ij} \le T^{\text{rescue}}, \forall i \in S^{\text{Set}}, j \in S^{\text{Station}} \tag{18}$$

$$\sum_{j \in S^{\text{Station}}} (L_{ij} + \frac{1}{2} l_{ij}^{\max}) p_j x_{ij} \le Q^{\text{Workload}}, \forall i \in S^{\text{Set}} \tag{19}$$

$$x_{ij}, y_i \in \{0,1\}, \forall i \in S^{\text{Set}}, j \in S^{\text{Station}} \tag{20}$$

The LCHSE has two objective functions, equation (11) is to maximize the total rescue distances of hot standby EMUs, and equation (12) is to minimize the number of rescue stations. The fewer rescue stations there are, the fewer hot standby EMUs there are. And the cost of producing hot standby EMUs is also relatively low. Constraint (14) shows that if station $k$ has been equipped with a hot standby EMU ( $k \in S^{\text{Depot}}$ ), it must be a rescue station. Constraint (15) is a logical constraint, if station $i$ is responsible for the rescue task of station $j$, it must be equipped with a hot standby EMU. The other constraints in LCHSE are the same as those in CMHSE.

To solve the model, we improve the objective functions of LCHSE. Because $Z_1 \geq 0$ and $Z_2 \geq 0$, there is:

$$\max\ Z_1 = \sum_{i \in S^{\text{Set}}} \sum_{j \in S^{\text{Station}}} (L_{ij} + \frac{1}{2} l_{ij}^{\max}) x_{ij} \Rightarrow \min\ \text{-}Z_1 = -\sum_{i \in S^{\text{Set}}} \sum_{j \in S^{\text{Station}}} (L_{ij} + \frac{1}{2} l_{ij}^{\max}) x_{ij} \tag{21}$$

In this way, the objective function of LCHSE can be improved as:

$$\min\ Z = -\beta \sum_{i \in S^{\text{Set}}} \sum_{j \in S^{\text{Station}}} (L_{ij} + \frac{1}{2} l_{ij}^{\max}) x_{ij} + \sum_{k \in S^{\text{Set}}} c_k y_k \tag{22}$$

where $\beta$ is the conversion coefficient that converts distance into cost.

## 5. Computational experiments

### 5.1 A case study of CMHSE

(1) Basic data

To verify the correctness of CMHSE, we conduct numerical analysis using the high-speed railway network shown in Figure 3. This network consists of 16 stations, designated as 1, 2, ..., 16, respectively. The numerical values shown on the lines connecting adjacent stations represent distances, measured in kilometers, for example, the distance between station 1 and station 2 is 197 km. Furthermore, the rescue stations, which are equipped with hot standby EMUs, are 5, 7, 9, 15 (green solid circle).

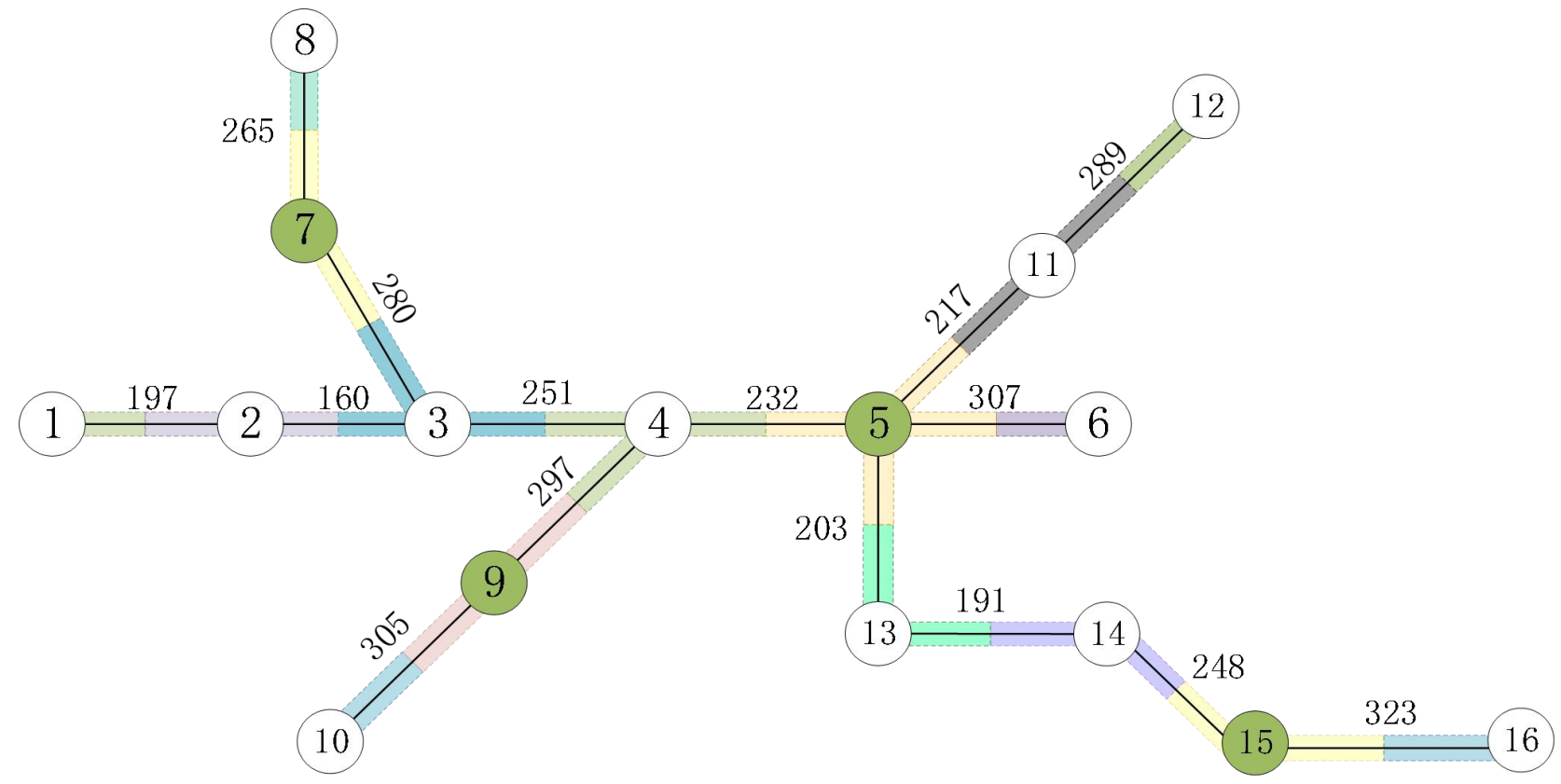

**Figure 3.** A high-speed railway network

Based on the 2022 China Railway Statistics Bulletin [30], the total distance of the high-speed railway network in China is 42000 kilometers, and the number of EMUs is 4194. The values of $\alpha$ and $\gamma$ are 0.025 and 0.8, respectively. The average speed of a hot standby EMU ($v$) is 300 km/h, and the value of $L^{\text{rescue}}$ is 800 km. Referring to the research by Xu et al. [31], the probability of accidents at a station ($p_j$) is 0.02. Additionally, the value of $T^{\text{rescue}}$ is 2.5 hours.

(2) Calculation results

The CMHSE is linear, and can be solved by commercial solvers such as Gurobi and CPLEX. In this study, we use Python 3.7 to call Gurobi 9.5.2 to solve CMHSE. The experiment is performed on a 3.10 GHz Intel (R) Core (TM) i5-7276U CPU computer with 4.0 GB of RAM. In the result, the objective function is 3963 km, and the variables with a value of 1 include: $x_{5,4}$, $x_{5,5}$ $x_{5,6}$, $x_{5,11}$, $x_{5,12}$, $x_{5,13}$, $x_{7,1}$, $x_{7,2}$, $x_{7,3}$, $x_{7,7}$, $x_{7,8}$, $x_{9,9}$, $x_{9,10}$, $x_{15,14}$, $x_{15,15}$, $x_{15,16}$.

Taking $x_{5,4}$ as an example, its value is 1, which means rescue station 5 is responsible for the rescue task of station 4. It is important to note that when a station is equipped with a hot standby EMU, it of course carries out the rescue task of itself, for example, $x_{5,5}=1$. Figure 4 shows the results of the experiment, the dashed lines with arrows indicate the coverage areas of the rescue stations, for example, rescue station 7 is responsible for the rescue tasks of stations 1, 2, 3, 7 and 8.

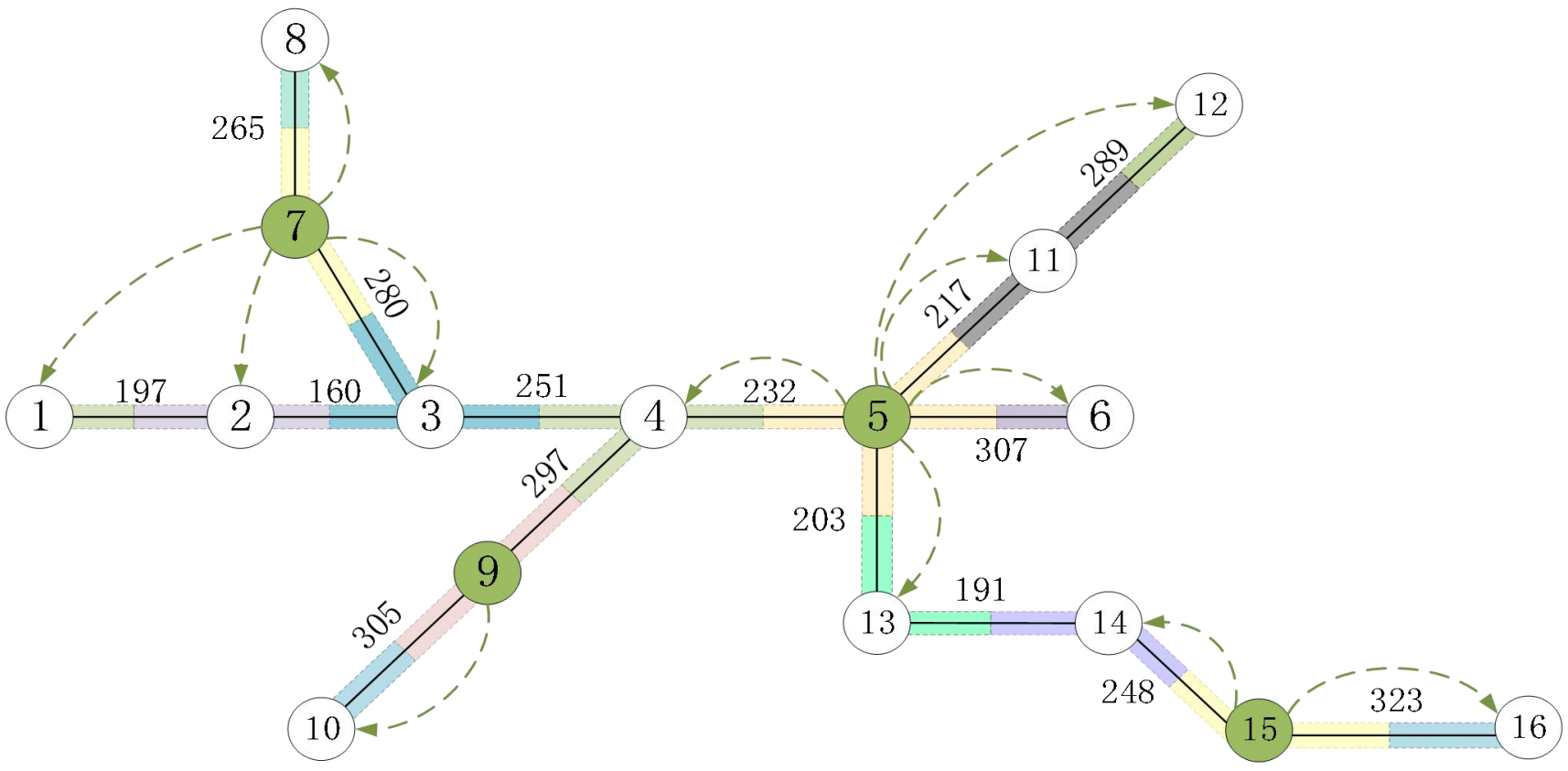

**Figure 4.** The coverage area of the rescue stations

In figure 4, we can also find that the result of CMHSE conforms to the principle of proximity in a physical network, while meeting the limitations of maximum rescue distance and maximum allowable rescue time.

### 5.2 A case study of LCHSE

(1) Basic data

We perform a computational experiment using the high-speed railway network shown in Figure 5. Notably, Figure 5 is an extension of the railway network shown in Figure 4, simulating an expansion in the high-speed railway infrastructure. There are 32 stations in Figure 5, and stations 5, 7, 9, 15 (green solid circle) are pre-given rescue stations which have been equipped with hot standby EMUs. Additionally, stations 3, 10, 11, 13, 16, 18, 22, and 26 (represented as orange solid circles) are potential rescue stations requiring optimization.

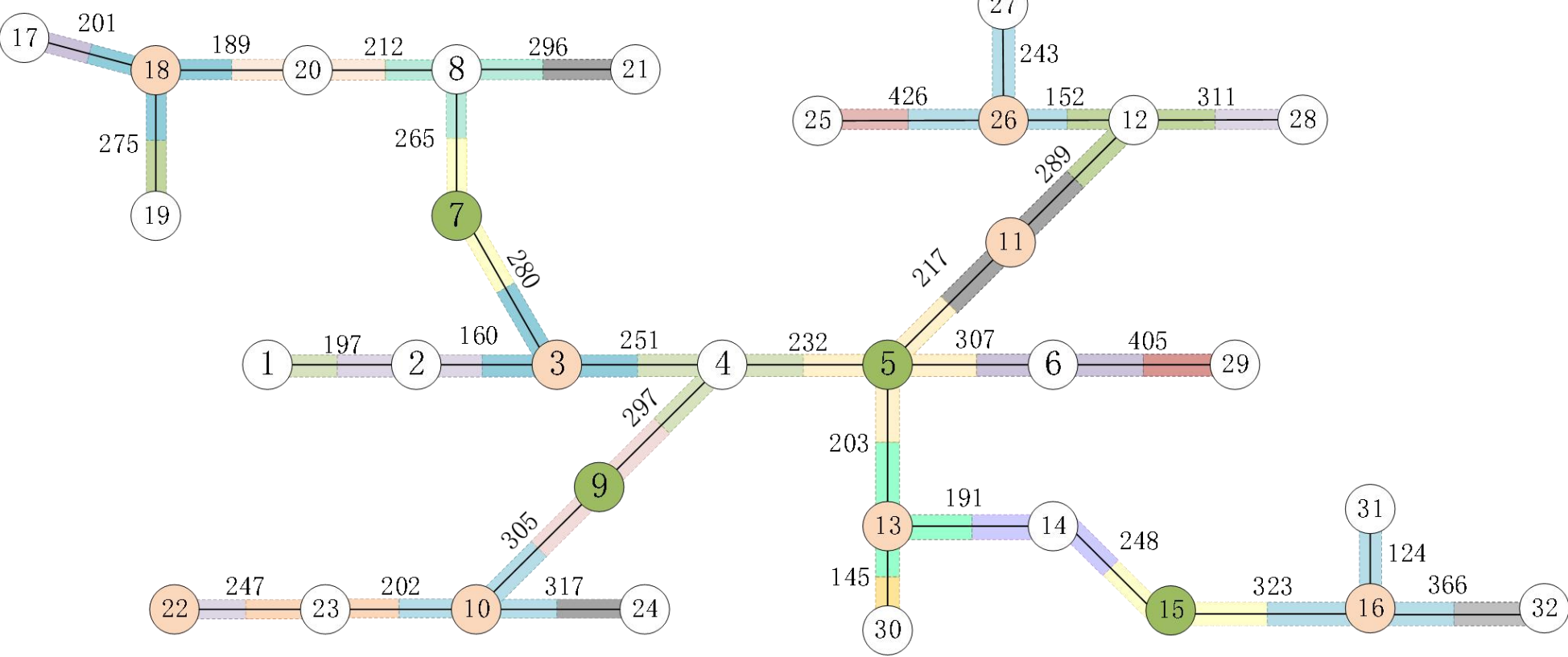

**Figure 5.** An expanded high-speed railway network

The cost of equipping a hot standby EMU at a station is 600000 yuan (RMB) and the value of $\beta$ is 200. All other parameters of LCHSE are the same as those in CMHSE. We also use Python 3.7 to call Gurobi 9.5.2 to solve LCHSE, and the experimental environment remains unchanged.

(2) Calculation results

In the results, the objective function of LCHSE is 149400 yuan, and rescue stations are 5, 7, 9, 15, 18, 26. Compared to the results of CMHSE, rescue stations 18 and 26 are newly added. The location and coverage area of these rescue stations in the high-speed railway network can be seen in Figure 6.

As we can see from figure 6, station 5 has the maximum workload, it carries out the rescue tasks for stations 4, 5, 6, 11, 13, 29, 30. And station 18 has the minimum workload, which just carries out the rescue tasks for three stations. But considering the high-speed railway network's expansion, the workload of rescue station 18 will also increase.

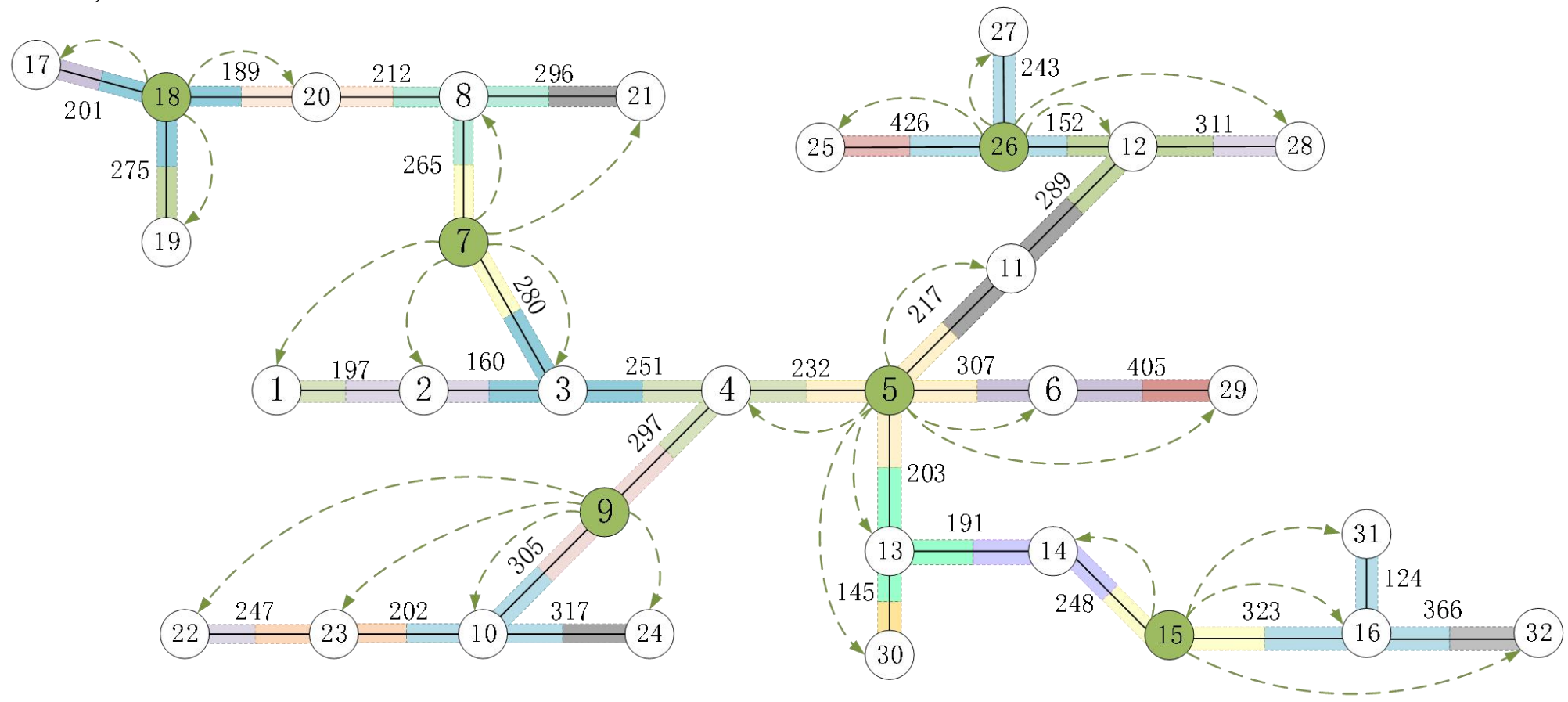

**Figure 6.** The results of LCHSE

## 6. Conclusions

This paper studies the location and coverage area of hot standby EMU, by using a merge strategy to divide the high-speed railway network into points, the problem is transformed into a "point-to-point" problem. At the end of the paper, the effectiveness of the models established in this paper is verified by numerical experiments.

Our optimization models can provide a scientific, reasonable and economic decision-making basis for railway company management departments, when planning and selecting rescue stations for hot standby EMUs. It can achieve the minimum number of hot standby EMUs covering the entire high-speed railway network, subsequently reducing the cost of producing hot standby EMUs.

In future research, our focus will shift to optimizing the location of EMU maintenance depots. Anticipating a surge in the number of EMUs, we foresee an impending peak period in EMU maintenance demand.

### Acknowledgments

The research was supported by the National Natural Science Foundation of China (U2268207).